\documentclass[12pt]{article}%
\usepackage{amsmath}
\usepackage{graphicx}%
\usepackage{amsfonts}%
\usepackage{amssymb}
\begin{document}

\begin{center}
{\LARGE Quantizing Dirac and Nambu Brackets\ }\footnote{Talk given by the
first author at the Coral Gables Conference 11-14 December 2002.}\bigskip

{\large Thomas Curtright}$^{\S}${\large \ and Cosmas Zachos}$^{\natural}$\bigskip

$^{\S}${Department of Physics, University of Miami}

{{Coral Gables, Florida 33124-8046, USA \textsl{curtright@physics.miami.edu}}}

$^{\natural}${High Energy Physics Division, Argonne National Laboratory}

{{Argonne, IL 60439-4815, USA \textsl{zachos@hep.anl.gov} } }\bigskip
\end{center}

\begin{quotation}
\noindent\textbf{Abstract.} \ We relate classical and quantum Dirac and Nambu
brackets. \ At the classical level, we use the relations between the two
brackets to gain some insight into the Jacobi identity for Dirac brackets,
among other things. \ At the quantum level, we suggest that the Nambu bracket
is the preferred method for introducing constraints, although at the expense
of some unorthodox behavior, which we describe in detail.\bigskip
\end{quotation}

\section{Introduction}

We have recently devoted some time to understand the quantization of Nambu
brackets (NBs) using non-Abelian methods\footnote{Our list of references here
will be abbreviated. \ A more complete set of references is given in the
correlated talk by C Zachos, also contributed to these Proceedings \cite{ZC}.}
\cite{CQNM}. \ Indirectly, our results are also a quantization of Dirac
brackets (DBs) because the two types of brackets are related, as we will
explain. \ Since it is the centenary of Dirac's birth, and since Dirac did
have an affiliation with the University of Miami, this Conference series, and
Behram Kursunoglu, it seemed appropriate to contribute something on this
subject to these Proceedings at this time. \ 

\section{Classical Theory}

\paragraph{\underline{Poisson and Dirac brackets}}

The ordinary Poisson bracket (PB) is the antisymmetric bilinear formed from
any pair of phase-space functions $A\left(  x,p\right)  ,\;B\left(
x,p\right)  $:%
\begin{equation}
\left\{  A,B\right\}  _{{\footnotesize PB}}=\sum_{i}\frac{\partial\left(
A,B\right)  }{\partial\left(  x_{i},p_{i}\right)  }=-\left\{  B,A\right\}
_{{\footnotesize PB}}\;.
\end{equation}
For a particle whose motion is constrained, however, this is not the most
appropriate tool to take apart and understand the dynamics. \ The Dirac
bracket is the preferred bilinear in the constrained case. \ It is defined for
any \textit{even} number of constraints. \ First compute the antisymmetric
matrix formed from PBs between pairs of phase-space constraints $C_{i}\left(
x,p\right)  $,%
\begin{equation}
A_{ij}=\left\{  C_{i},C_{j}\right\}  _{{\footnotesize PB}}\;. \label{CMatrix}%
\end{equation}
Then compute it's inverse, $A_{ij}^{-1}$, assuming the latter exists. \ 

The Dirac bracket \cite{Dirac} is then defined in terms of PBs as%
\begin{equation}
\left\{  f,g\right\}  _{{\footnotesize DB}}=\left\{  f,g\right\}
_{{\footnotesize PB}}-\sum_{i,j}\left\{  f,C_{i}\right\}  _{{\footnotesize PB}%
}A_{ij}^{-1}\left\{  C_{j},g\right\}  _{{\footnotesize PB}}=-\left\{
g,f\right\}  _{{\footnotesize DB}}\;. \label{DBDefn}%
\end{equation}
By construction, the DB of any individual constraint, among those incorporated
into the DB definition, with \emph{any} function $f$ on the phase-space, will
vanish identically.%
\begin{equation}
\left\{  C_{k},f\right\}  _{{\footnotesize DB}}=\left\{  C_{k},f\right\}
_{{\footnotesize PB}}-\sum_{i,j}\left\{  C_{k},C_{i}\right\}
_{{\footnotesize PB}}A_{ij}^{-1}\left\{  C_{j},f\right\}  _{{\footnotesize PB}%
}=\left\{  C_{k},f\right\}  _{{\footnotesize PB}}-\left\{  C_{k},f\right\}
_{{\footnotesize PB}}=0\;.
\end{equation}

\paragraph{\underline{The Jacobi Identity}}

A nontrivial property is the Jacobi identity (JI) for the Dirac bracket: \
\begin{equation}
0=\left\{  f,\left\{  g,h\right\}  _{{\footnotesize DB}}\right\}
_{{\footnotesize DB}}+\left\{  g,\left\{  h,f\right\}  _{{\footnotesize DB}%
}\right\}  _{{\footnotesize DB}}+\left\{  h,\left\{  f,g\right\}
_{{\footnotesize DB}}\right\}  _{{\footnotesize DB}}\;. \label{DBJI}%
\end{equation}
This is not so easily demonstrated by direct calculation. \ It is surprisingly
difficult (see p 42, \textit{L on QM}, \cite{Dirac}) to establish the JI for
the DB without relating it to something else. \ We will need
\begin{equation}
\delta A^{-1}=-A^{-1}\left(  \delta A\right)  A^{-1}\;,
\label{DerivationOfInverse}%
\end{equation}
as is valid for any derivation $\delta$, as well as the JI for the Poisson
bracket. \ The Jacobi identity is itself a property very similar in form to
the derivation property, only it applies to the action of one bracket
functional on another. So, with%
\begin{equation}
\delta_{f}\,g\equiv\left\{  f,g\right\}  _{{\footnotesize PB}}\;,
\end{equation}
not only do we have%
\begin{equation}
\delta_{f}\,\left(  gh\right)  \equiv\left\{  f,gh\right\}
_{{\footnotesize PB}}=\left(  \delta_{f}\,g\right)  h+g\left(  \delta
_{f}\,h\right)  \;,
\end{equation}
but we also have%
\begin{equation}
\delta_{f}\,\left(  \left\{  g,h\right\}  _{{\footnotesize PB}}\right)
\equiv\left\{  f,\left\{  g,h\right\}  _{{\footnotesize PB}}\right\}
_{{\footnotesize PB}}=\left\{  \left(  \delta_{f}\,g\right)  ,h\right\}
_{{\footnotesize PB}}+\left\{  g,\left(  \delta_{f}\,h\right)  \right\}
_{{\footnotesize PB}}\;.
\end{equation}
That is to say%
\begin{equation}
0=\left\{  f,\left\{  g,h\right\}  _{{\footnotesize PB}}\right\}
_{{\footnotesize PB}}+\left\{  g,\left\{  h,f\right\}  _{{\footnotesize PB}%
}\right\}  _{{\footnotesize PB}}+\left\{  h,\left\{  f,g\right\}
_{{\footnotesize PB}}\right\}  _{{\footnotesize PB}}\;. \label{PBJI}%
\end{equation}
Now as a consequence of the PB being a derivation, it is trivially true, for
commuting quantities, that
\begin{equation}
\left\{  f,gh\right\}  _{{\footnotesize DB}}=\left\{  f,gh\right\}
_{{\footnotesize PB}}-\sum_{i,j}\left\{  f,C_{i}\right\}  _{{\footnotesize PB}%
}A_{ij}^{-1}\left\{  C_{j},gh\right\}  _{{\footnotesize PB}}=\left\{
f,g\right\}  _{{\footnotesize DB}}h+g\left\{  f,h\right\}
_{{\footnotesize DB}}\;,
\end{equation}
so the DB is also a derivation. \ But it is not so obviously true that the DB
acts in an analogous fashion on other DBs:%
\begin{equation}
\left\{  f,\left\{  g,h\right\}  \right\}  _{{\footnotesize DB}}=\left\{
\left\{  f,g\right\}  _{{\footnotesize DB}},h\right\}  _{{\footnotesize DB}%
}+\left\{  g,\left\{  f,h\right\}  _{{\footnotesize DB}}\right\}
_{{\footnotesize DB}}\;. \label{DBDerivation}%
\end{equation}
This is exactly what we want to establish to prove the JI for DBs. \ The terms
that get in the way involve the DB acting on the constraint functionals:
\ $\left\{  \cdot,C\right\}  _{PB}$ and $A^{-1}$. \ Fortunately, not only are
the constraints themselves invariant under the action of the DB, but so are
these functionals, at least to a sufficient degree (as explained below).
\ Thus the Jacobi identity ultimately holds.

However, just for extra fun, let us use only the derivation property of the
Poisson bracket, and \emph{not perform any re-ordering of terms} (in
anticipation of quantization, so that all products could be non-commutative
$\star$ products \cite{ZC}). By direct calculation, we then obtain a slew of
terms (arranged in cyclic triples):%
\[
\left\{  f,\left\{  g,h\right\}  _{{\footnotesize DB}}\right\}
_{{\footnotesize DB}}+\left\{  g,\left\{  h,f\right\}  _{{\footnotesize DB}%
}\right\}  _{{\footnotesize DB}}+\left\{  h,\left\{  f,g\right\}
_{{\footnotesize DB}}\right\}  _{{\footnotesize DB}}=
\]
\begin{subequations}
\label{DBJITest}%
\begin{gather}
\left\{  f,\left\{  g,h\right\}  \right\}  +\left\{  g,\left\{  h,f\right\}
\right\}  +\left\{  h,\left\{  f,g\right\}  \right\} \\
\nonumber\\
{\small +}\left\{  C_{i},f\right\}  {\small A}_{im}^{-1}\left\{
A_{mn},h\right\}  {\small A}_{nj}^{-1}\left\{  C_{j},g\right\}  {\small -}%
\tfrac{1}{2}\left\{  C_{i},f\right\}  {\small A}_{im}^{-1}\left\{
C_{j},g\right\}  {\small A}_{nj}^{-1}\left\{  A_{mn},h\right\} \nonumber\\
{\small -}\tfrac{1}{2}\left\{  C_{j},g\right\}  {\small A}_{nj}^{-1}\left\{
A_{mn},h\right\}  {\small A}_{im}^{-1}\left\{  C_{i},f\right\} \\
{\small +}\left\{  C_{i},g\right\}  {\small A}_{im}^{-1}\left\{
A_{mn},f\right\}  {\small A}_{nj}^{-1}\left\{  C_{j},h\right\}  {\small -}%
\tfrac{1}{2}\left\{  C_{i},g\right\}  {\small A}_{im}^{-1}\left\{
C_{j},h\right\}  {\small A}_{nj}^{-1}\left\{  A_{mn},f\right\} \nonumber\\
{\small -}\tfrac{1}{2}\left\{  C_{j},h\right\}  {\small A}_{nj}^{-1}\left\{
A_{mn},f\right\}  {\small A}_{im}^{-1}\left\{  C_{i},g\right\} \\
{\small +}\left\{  C_{i},h\right\}  {\small A}_{im}^{-1}\left\{
A_{mn},g\right\}  {\small A}_{nj}^{-1}\left\{  C_{j},f\right\}  {\small -}%
\tfrac{1}{2}\left\{  C_{i},h\right\}  {\small A}_{im}^{-1}\left\{
C_{j},f\right\}  {\small A}_{nj}^{-1}\left\{  A_{mn},g\right\} \nonumber\\
{\small -}\tfrac{1}{2}\left\{  C_{j},f\right\}  {\small A}_{nj}^{-1}\left\{
A_{mn},g\right\}  {\small A}_{im}^{-1}\left\{  C_{i},h\right\} \\
\nonumber\\
+\left\{  C_{i},f\right\}  A_{im}^{-1}\left\{  C_{j},g\right\}  A_{jn}%
^{-1}S_{mn}\left(  h\right)  -\left\{  C_{j},g\right\}  A_{jn}^{-1}%
S_{mn}\left(  h\right)  A_{im}^{-1}\left\{  C_{i},f\right\} \\
+\left\{  C_{i},g\right\}  A_{im}^{-1}\left\{  C_{j},h\right\}  A_{jn}%
^{-1}S_{mn}\left(  f\right)  -\left\{  C_{j},h\right\}  A_{jn}^{-1}%
S_{mn}\left(  f\right)  A_{im}^{-1}\left\{  C_{i},g\right\} \\
+\left\{  C_{i},h\right\}  A_{im}^{-1}\left\{  C_{j},f\right\}  A_{jn}%
^{-1}S_{mn}\left(  g\right)  -\left\{  C_{j},f\right\}  A_{jn}^{-1}%
S_{mn}\left(  g\right)  A_{im}^{-1}\left\{  C_{i},h\right\} \\
\nonumber\\
+\left\{  C_{i},f\right\}  A_{ij}^{-1}\left\{  C_{j},\left\{  g,h\right\}
\right\}  +\left\{  C_{i},f\right\}  A_{ij}^{-1}\left\{  h,\left\{
C_{j},g\right\}  \right\}  +\left\{  g,\left\{  C_{i},h\right\}  \right\}
A_{ij}^{-1}\left\{  C_{j},f\right\} \\
+\left\{  C_{i},g\right\}  A_{ij}^{-1}\left\{  C_{j},\left\{  h,f\right\}
\right\}  +\left\{  C_{i},g\right\}  A_{ij}^{-1}\left\{  f,\left\{
C_{j},h\right\}  \right\}  +\left\{  h,\left\{  C_{i},f\right\}  \right\}
A_{ij}^{-1}\left\{  C_{j},g\right\} \\
+\left\{  C_{i},h\right\}  A_{ij}^{-1}\left\{  C_{j},\left\{  f,g\right\}
\right\}  +\left\{  C_{i},h\right\}  A_{ij}^{-1}\left\{  g,\left\{
C_{j},f\right\}  \right\}  +\left\{  f,\left\{  C_{i},g\right\}  \right\}
A_{ij}^{-1}\left\{  C_{j},h\right\} \\
\nonumber\\
+\left\{  C_{i},f\right\}  A_{ij}^{-1}\left\{  C_{k},g\right\}  \left\{
C_{j},A_{kl}^{-1}\right\}  \left\{  C_{l},h\right\}  +\left\{  C_{i}%
,g\right\}  A_{ij}^{-1}\left\{  C_{k},h\right\}  \left\{  C_{j},A_{kl}%
^{-1}\right\}  \left\{  C_{l},f\right\} \nonumber\\
+\left\{  C_{i},h\right\}  A_{ij}^{-1}\left\{  C_{k},f\right\}  \left\{
C_{j},A_{kl}^{-1}\right\}  \left\{  C_{l},g\right\}
\end{gather}
For purposes of the classical discussion, all the unlabeled brackets on the
RHS of (\ref{DBJITest}) are PBs. \ Here we have also defined
\end{subequations}
\begin{equation}
S_{jk}\left(  f\right)  =S_{kj}\left(  f\right)  =\tfrac{1}{2}\left\{
C_{j},\left\{  C_{k},f\right\}  \right\}  +\tfrac{1}{2}\left\{  C_{k},\left\{
C_{j},f\right\}  \right\}  \;,
\end{equation}
and similarly for $S_{jk}\left(  g\right)  $ and $S_{jk}\left(  h\right)  $.

Now, if all quantities commute, as is the case for classical PBs, then each
individually numbered line on the RHS of (\ref{DBJITest}) vanishes separately.
\ The only line presenting any issue is perhaps the last one. \ However, for
commuting quantities it is just%
\begin{gather}
\left\{  C_{i},f\right\}  \left\{  C_{k},g\right\}  \left\{  C_{l},h\right\}
\left(  A_{ij}^{-1}\left\{  C_{j},A_{kl}^{-1}\right\}  +A_{kj}^{-1}\left\{
C_{j},A_{li}^{-1}\right\}  +A_{lj}^{-1}\left\{  C_{j},A_{ik}^{-1}\right\}
\right) \nonumber\\
=-\left\{  C_{i},f\right\}  \left\{  C_{k},g\right\}  \left\{  C_{l}%
,h\right\}  A_{ij}^{-1}A_{km}^{-1}A_{nl}^{-1}\times\nonumber\\
\times\left(  \left\{  C_{j},A_{mn}\right\}  +\left\{  C_{m},A_{nj}\right\}
+\left\{  C_{n},A_{jm}\right\}  \right)  \;,
\end{gather}
where in the last expression, we used (\ref{DerivationOfInverse}) written as
$\left\{  C_{j},A_{kl}^{-1}\right\}  =-A_{km}^{-1}\left\{  C_{j}%
,A_{mn}\right\}  A_{nl}^{-1}$, etc., and the antisymmetry of $A^{-1}$. \ But
now, from the definition (\ref{CMatrix}),%
\begin{equation}
\left\{  C_{j},A_{mn}\right\}  _{PB}+\left\{  C_{m},A_{nj}\right\}
_{PB}+\left\{  C_{n},A_{jm}\right\}  _{PB}=0
\end{equation}
is just the JI for PBs. \ Thus when all quantities commute, the DB JI is
established. \ But when all quantities do \emph{not} commute, the individual
lines on the RHS of (\ref{DBJITest}) do \emph{not} vanish, in general, nor
does their sum.

Now, clearly, what we have done is \emph{not quite correct} if things do not
commute. \ Our definition of the Dirac bracket $\left\{  f,g\right\}  _{DB}$
is not manifestly antisymmetric in $f$ and $g$ if the products involved in
$\left\{  f,C_{i}\right\}  A_{ij}^{-1}\left\{  C_{j},g\right\}  $ do not
commute. \ We should have at least defined%
\begin{equation}
\left\{  f,g\right\}  _{{\footnotesize DB}}=\left\{  f,g\right\}  -\tfrac
{1}{2}\left\{  f,C_{i}\right\}  A_{ij}^{-1}\left\{  C_{j},g\right\}
+\tfrac{1}{2}\left\{  g,C_{i}\right\}  A_{ij}^{-1}\left\{  C_{j},f\right\}
\;.
\end{equation}
Perhaps there is even a better definition. \ In any case, our previous answer
can be corrected just by antisymmetrizing the constraint terms with respect to
such interchanges. \ But this really does not make it any easier to see how
the terms on the RHS of (\ref{DBJITest}) can cancel, in general, when
non-commuting products are involved. \ To better understand how all this
machinery functions, it would be technically sweet to have another route to quantization.

\paragraph{\underline{Classical Nambu brackets}}

There is a multi-linear, totally antisymmetric bracket, introduced by Nambu
\cite{Nambu}. \ On the full phase-space for a particle with $N$ degrees of
freedom the highest rank, or \emph{maximal,} Nambu bracket (NB) is essentially
just the Jacobian
\begin{equation}
\left\{  A_{1},A_{2},\cdots,A_{2N}\right\}  _{{\footnotesize NB}%
}=\frac{\partial\left(  A_{1},A_{2},\cdots,A_{2N}\right)  }{\partial\left(
x_{1},p_{1},x_{2},p_{2},\cdots,x_{N},p_{N}\right)  }\;.
\end{equation}
Physically, this has a geometrical interpretation in terms of phase-space
gradients. \
\begin{center}
\includegraphics[
height=4.2237in,
width=6.2898in
]%
{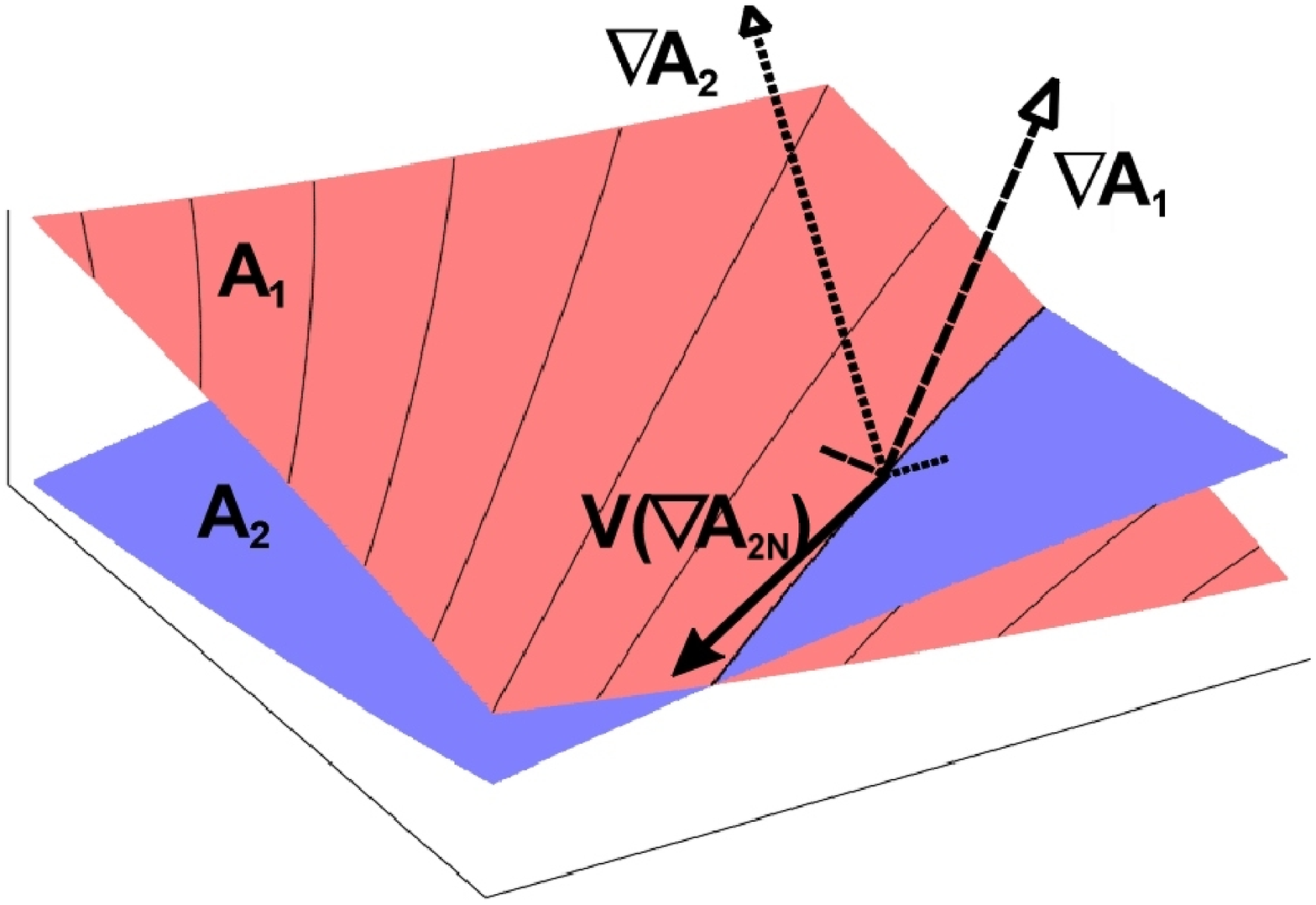}%
\end{center}
The surfaces illustrated in the Figure are isoclines for two different
phase-space functions, respectively $A_{1}$ and $A_{2}$. \ A particular
phase-space tangent $\mathsf{v}$ lies at the intersection of these two
surfaces. \ That local phase-space tangent at the point depicted is actually
given by the joint cross-product of all but one (e.g. $A_{2N}$) of the local
phase-space gradients of the $A$s contained within the NB. \ The complete,
maximal NB provides a projection onto $\mathsf{v}$ of the remaining
phase-space gradient (e.g. $\nabla A_{2N}$). \ Other possible intersections
along the $A_{1}$ surface are also shown as contours representing other values
for $A_{2}$, but the corresponding constant $A_{2}$ surfaces are not shown.

\paragraph{\underline{The Fundamental Identity}}

The classical NB obeys a simple combinatorial identity known as the
``fundamental identity'', or simply ``FI''. \ If you attempt to totally
antisymmetrize $2N+1$ commuting entities chosen from among $2N$ possibilities,
you obviously obtain zero. \ As a consequence,%
\begin{gather}
\left\{  \left\{  A_{1},\cdots,A_{2N}\right\}  _{{\footnotesize NB}}%
,B_{2},\cdots,B_{2N}\right\}  _{{\footnotesize NB}}=\left\{  \left\{
A_{1},B_{2},\cdots,B_{2N}\right\}  _{{\footnotesize NB}},A_{2},\cdots
,A_{2N}\right\}  _{{\footnotesize NB}}\nonumber\\
+\left\{  A_{1},\left\{  A_{2},B_{2},\cdots,B_{2N}\right\}
_{{\footnotesize NB}},\cdots,A_{2N}\right\}  _{{\footnotesize NB}}%
+\cdots+\left\{  A_{1},\cdots,\left\{  A_{2N},B_{2},\cdots,B_{2N}\right\}
_{{\footnotesize NB}}\right\}  _{{\footnotesize NB}}. \label{Fundamental}%
\end{gather}
This has the form needed to say that the action of one maximal NB functional
on another is similar to a derivation. \ That is, in an obvious notation,%
\begin{equation}
\delta_{NB}\left\{  A_{1},A_{2},\cdots,A_{2N}\right\}  _{{\footnotesize NB}%
}=\left\{  \delta_{NB}A_{1},\cdots,A_{2N}\right\}  _{{\footnotesize NB}%
}+\cdots+\left\{  A_{1},\cdots,\delta_{NB}A_{2N}\right\}  _{{\footnotesize NB}%
}\;.
\end{equation}
It is a bit misleading to call this particular identity ``fundamental'', since
the same relation holds whenever any NB is applied to $\left\{  A_{1}%
,A_{2},\cdots,A_{2N}\right\}  _{{\footnotesize NB}}$. \ In other words,
classically there are a set of such identities. \ We will discuss them all
below, after defining NBs with fewer entries. \ More importantly, however, it
is a less fortunate misnomer to call (\ref{Fundamental}) ``fundamental'' from
the standpoint of the quantum theory discussed below.

\paragraph{\underline{DBs as NBs}}

There are some elementary but important relationships between Dirac brackets
and Nambu brackets \cite{GN,CZ}.

For a particle with $N$ degrees of freedom but subject to $2n$ constraints,
consider a Nambu bracket involving any two functions $A,B$ of the dynamical
variables on the full $2N$ dimensional phase-space. \ Along with $A,B$ insert
into the bracket all of the constraints as well as $k=N-n-1$ factors of
$x_{i},p_{i}$ and sum from $1$ to $N$ all $k$ of the $i$'s, i.e. take $k$
\emph{symplectic traces}. \ This reduces the Nambu bracket to a Dirac bracket.
\ Thus%
\begin{equation}
\left\{  A,B\right\}  _{{\footnotesize DB}}\propto\sum_{i_{1},\cdots,i_{k}%
=1}^{N}\frac{\partial\left(  A,B,C_{1},C_{2},\cdots,C_{2n},x_{i_{1}},p_{i_{1}%
},\cdots,x_{i_{k}},p_{i_{k}}\right)  }{\partial\left(  x_{1},p_{1}%
,\cdots,x_{N},p_{N}\right)  }\;.
\end{equation}
with the proportionality to be determined (see (\ref{SymplecticTrace}) and
(\ref{DBasNB}) below). \ Hence the Dirac bracket always follows directly from
the Nambu bracket.

Any maximal even rank CNB can also be resolved into products of Poisson
brackets. \ For example, for systems with two degrees of freedom, $\left\{
A,B\right\}  _{{\footnotesize PB}}=\frac{\partial\left(  A,B\right)
}{\partial\left(  x_{1},p_{1}\right)  }+\frac{\partial\left(  A,B\right)
}{\partial\left(  x_{2},p_{2}\right)  }$, and the 4-bracket $\left\{
A,B,C,D\right\}  _{{\footnotesize NB}}\equiv\frac{\partial\left(
A,B,C,D\right)  }{\partial\left(  x_{1},p_{1},x_{2},p_{2}\right)  }$ resolves
as%
\begin{equation}
\left\{  A,B,C,D\right\}  _{{\footnotesize NB}}=\left\{  A,B\right\}
_{{\footnotesize PB}}\left\{  C,D\right\}  _{{\footnotesize PB}}-\left\{
A,C\right\}  _{{\footnotesize PB}}\left\{  B,D\right\}  _{{\footnotesize PB}%
}-\left\{  A,D\right\}  _{{\footnotesize PB}}\left\{  C,B\right\}
_{{\footnotesize PB}}\;,
\end{equation}
in comportance with full antisymmetry under permutations of $A,B,C,$ and $D$.
\ The general result for maximal rank $2N$ brackets for systems with a
$2N$-dimensional phase-space is%
\begin{equation}
\left\{  A_{1},\cdots,A_{2N}\right\}  _{{\footnotesize NB}}=\sum
_{\substack{\text{all }\left(  2N\right)  !\text{\ perms } \\\left\{
\sigma_{1},\sigma_{2},\cdots,\sigma_{2N}\right\}  \\\text{of the indices }
\\\left\{  1,2,\cdots,2N\right\}  }}\frac{\operatorname{sgn}\left(
\sigma\right)  }{2^{N}N!}\,\left\{  A_{\sigma_{1}},A_{\sigma_{2}}\right\}
_{{\footnotesize PB}}\cdots\left\{  A_{\sigma_{2N-1}},A_{\sigma_{2N}}\right\}
_{{\footnotesize PB}}\;, \label{PBResolution}%
\end{equation}
where $\operatorname{sgn}\left(  \sigma\right)  =\left(  -1\right)
^{\pi\left(  \sigma\right)  }$ with $\pi\left(  \sigma\right)  $\ the parity
of the permutation $\left\{  \sigma_{1},\sigma_{2},\cdots,\sigma_{2N}\right\}
$. \ The sum only gives $\left(  2N-1\right)  !!=\left(  2N\right)  !/\left(
2^{N}N!\right)  $ distinct products of PBs on the RHS, not $\left(  2N\right)
!$ \ Each such distinct product appears with net coefficient $\pm1$. \ 

The proof of the relation (\ref{PBResolution}) is remarkably elementary.
\ Both left- and right-hand sides of the expression are sums of $2N$-th degree
monomials linear in the $2N$ first-order partial derivatives of each of the
$A$s. Both sides are totally antisymmetric under permutations of the $A$s.
\ Hence, both sides are also totally antisymmetric under interchanges of
partial derivatives. \ Thus, the two sides must be proportional. \ The only
issue left is the constant of proportionality. \ This is easily determined to
be $1$ just by comparing the coefficients of any given term appearing on both
sides of the equation, e.g. $\partial_{x_{1}}A_{1}\partial_{p_{1}}A_{2}%
\cdots\partial_{x_{N}}A_{2N-1}\partial_{p_{N}}A_{2N}$.

This is essentially a special case of Laplace's theorem on the general minor
expansions of determinants, although it must be said that we have never seen
it written, let alone used, in exactly this form, either in treatises on
determinants or in textbooks on classical mechanics.

For similar relations to hold for sub-maximal even rank Nambu brackets, these
must first be defined. \ It is easiest to just \emph{define} sub-maximal even
rank CNBs by their Poisson bracket resolutions:%
\begin{equation}
\left\{  A_{1},\cdots,A_{2n}\right\}  _{{\footnotesize NB}}=\sum
_{\substack{\left(  2n\right)  !\text{\ perms }\sigma}%
}\frac{\operatorname{sgn}\left(  \sigma\right)  }{2^{n}n!}\,\left\{
A_{\sigma_{1}},A_{\sigma_{2}}\right\}  _{{\footnotesize PB}}\cdots\left\{
A_{\sigma_{2n-1}},A_{\sigma_{2n}}\right\}  _{{\footnotesize PB}}\;,
\end{equation}
only here we allow $n<N$. \ So defined, these sub-maximal CNBs enter in
further recursive expressions. \ For example, for systems with three or more
degrees of freedom, $\left\{  A,B\right\}  _{{\footnotesize PB}}%
=\frac{\partial\left(  A,B\right)  }{\partial\left(  x_{1},p_{1}\right)
}+\frac{\partial\left(  A,B\right)  }{\partial\left(  x_{2},p_{2}\right)
}+\frac{\partial\left(  A,B\right)  }{\partial\left(  x_{3},p_{3}\right)
}+\cdots$, and a general 6-bracket expression resolves as
\begin{align}
&  \left\{  A_{1},A_{2},A_{3},A_{4},A_{5},A_{6}\right\}  _{{\footnotesize NB}%
}\nonumber\\
&  =\left\{  A_{1},A_{2}\right\}  _{{\footnotesize PB}}\left\{  A_{3}%
,A_{4},A_{5},A_{6}\right\}  _{{\footnotesize NB}}-\left\{  A_{1}%
,A_{3}\right\}  _{{\footnotesize PB}}\left\{  A_{2},A_{4},A_{5},A_{6}\right\}
_{{\footnotesize NB}}\nonumber\\
&  +\left\{  A_{1},A_{4}\right\}  _{{\footnotesize PB}}\left\{  A_{2}%
,A_{3},A_{5},A_{6}\right\}  _{{\footnotesize NB}}-\left\{  A_{1}%
,A_{5}\right\}  _{{\footnotesize PB}}\left\{  A_{2},A_{3},A_{4},A_{6}\right\}
_{{\footnotesize NB}}\nonumber\\
&  +\left\{  A_{1},A_{6}\right\}  _{{\footnotesize PB}}\left\{  A_{2}%
,A_{3},A_{4},A_{5}\right\}  _{{\footnotesize NB}}\;,
\end{align}
with the 4-brackets resolvable into PBs as above. \ Sub-maximal brackets
defined in this way are the same as those obtained by taking symplectic traces
of maximal brackets.%

\begin{equation}
\left\{  A_{1},\cdots,A_{2n}\right\}  _{{\footnotesize NB}}=\frac{1}{\left(
N-n\right)  !}\sum_{i_{1},\cdots,i_{N-n}=1}^{N}\frac{\partial\left(
A_{1},\cdots,A_{2n},x_{i_{1}},p_{i_{1}},\cdots,x_{i_{N-n}},p_{i_{N-n}}\right)
}{\partial\left(  x_{1},p_{1},\cdots,x_{N},p_{N}\right)  }\;.
\label{SymplecticTrace}%
\end{equation}
This permits the building-up of higher even rank brackets proceeding from
initial PBs involving all degrees of freedom. \ The general recursion relation
with this $2n=2+\left(  2n-2\right)  $ form is%
\begin{align}
\left\{  A_{1},\cdots,A_{2n}\right\}  _{{\footnotesize NB}}  &  =\left\{
A_{1},A_{2}\right\}  _{{\footnotesize PB}}\left\{  A_{3},\cdots,A_{2n}%
\right\}  _{{\footnotesize NB}}\nonumber\\
&  +\sum_{j=3}^{2n-1}\left(  -1\right)  ^{j}\left\{  A_{1},A_{j}\right\}
_{{\footnotesize PB}}\left\{  A_{2},\cdots,A_{j-1},A_{j+1},\cdots
,A_{2n}\right\}  _{{\footnotesize NB}}\nonumber\\
&  +\left\{  A_{1},A_{2n}\right\}  _{{\footnotesize PB}}\left\{  A_{2}%
,\cdots,A_{2n-1}\right\}  _{{\footnotesize NB}}\;, \label{NBRecursion}%
\end{align}
and features $2n-1$ terms on the RHS. \ (Of course, one can permute the
subscripts to get equivalent forms for the RHS.) \ This recursive result is
equivalent to taking the PB resolution as a definition for $2n<2N$ elements,
as can be seen by substituting the PB resolutions of the $\left(  2n-2\right)
$-brackets on the RHS. \ Similar relations obtain\ when the $2n$ elements in
the CNB are partitioned into sets of $\left(  2n-2k\right)  $ and $2k$
elements, with suitable antisymmetrization with respect to exchanges between
the two sets.

Having defined CNBs with fewer than $2N$ entries, we present the sub-maximal
extensions of the so-called fundamental identity. \ These new identities can
be proven either by using the PB resolutions, or by taking symplectic traces
of (\ref{Fundamental}). \ For arbitrary strings of elements, $\mathbf{A}%
=A_{1},\cdots,A_{n}$ and $\mathbf{B=}B_{1},\cdots,B_{k}$, with $n$ even and
$k$ odd, and for any additional phase-space ``weight'' $V$, it follows that%
\begin{align}
&  \left\{  \mathbf{B},V\left\{  \mathbf{A}\right\}  _{{\footnotesize NB}%
}\right\}  _{{\footnotesize NB}}-\left\{  V\left\{  \mathbf{B},A_{1}\right\}
_{{\footnotesize NB}},A_{2},\cdots,A_{n}\right\}  _{{\footnotesize NB}}%
-\cdots-\left\{  A_{1},A_{2},\cdots,V\left\{  \mathbf{B},A_{n}\right\}
_{{\footnotesize NB}}\right\}  _{{\footnotesize NB}}\nonumber\\
&  =\left\{  B_{1},V\left\{  B_{2},B_{3},\cdots,B_{k}\right\}
_{{\footnotesize NB}},\mathbf{A}\right\}  _{{\footnotesize NB}}-\cdots
+\left\{  B_{k},V\left\{  B_{1},B_{2},\cdots,B_{k-1}\right\}
_{{\footnotesize NB}},\mathbf{A}\right\}  _{{\footnotesize NB}}\;,
\label{Fundamentalest}%
\end{align}
in an obvious notation: \ $\left\{  \mathbf{B},V\left\{  \mathbf{A}\right\}
_{{\footnotesize NB}}\right\}  _{{\footnotesize NB}}=\left\{  B_{1}%
,\cdots,B_{k},V\left\{  A_{1},\cdots,A_{n}\right\}  _{{\footnotesize NB}%
}\right\}  _{{\footnotesize NB}}$, etc. \ When the $\mathbf{A}$ string is
maximal, i.e. $n=2N$, the RHS of (\ref{Fundamentalest}) vanishes identically,
for any odd value of $k$. \ (The RHS also vanishes identically when $k=1,$ for
any even $n,$ if $V$ is a numerical constant.) \ We may use
(\ref{Fundamentalest})\ to prove the DB JI, but first we need to express DBs
precisely as NBs.

The PB resolutions lead to a precise, useful relation between Nambu and Dirac
brackets, valid for any even number of constraints. \ Take $A_{1}%
=f,\;A_{2}=g,\;A_{3}=C_{1},\cdots,$ then for a $2n$-bracket containing $f$,
$g$, and $\left(  2n-2\right)  $ constraints, we first extract $f$ into a PB
in all possible ways.%
\begin{gather}
\left\{  f,g,C_{1},C_{2},\cdots\right\}  _{{\footnotesize NB}}\nonumber\\
=\left\{  f,g\right\}  _{{\footnotesize PB}}\left\{  C_{1},C_{2}%
,\cdots\right\}  _{{\footnotesize NB}}+\sum_{j=1}^{2n-2}\left(  -1\right)
^{j}\left\{  f,C_{j}\right\}  _{{\footnotesize PB}}\left\{  g,C_{1}%
,\cdots,C_{j-1},C_{j+1},\cdots\right\}  _{{\footnotesize NB}}\;.
\end{gather}
Likewise extract $g$ as a second sum, taking care to note that $C_{j}$ has
already been removed from the remaining Nambu bracket. \ Thus%
\begin{gather}
\left\{  f,g,C_{1},C_{2},\cdots\right\}  _{{\footnotesize NB}}=\left\{
f,g\right\}  _{{\footnotesize PB}}\left\{  C_{1},C_{2},\cdots\right\}
_{{\footnotesize NB}}\nonumber\\
+\sum_{j=2}^{2n-2}\left\{  f,C_{j-1}\right\}  _{{\footnotesize PB}}\left\{
C_{1},\cdots,C_{j-2},C_{j+1},\cdots,C_{2n-2}\right\}  _{{\footnotesize NB}%
}\left\{  C_{j},g\right\}  _{{\footnotesize PB}}\nonumber\\
-\sum_{j=2}^{2n-2}\left\{  f,C_{j}\right\}  _{{\footnotesize PB}}\left\{
C_{1},\cdots,C_{j-2},C_{j+1},\cdots,C_{2n-2}\right\}  _{{\footnotesize NB}%
}\left\{  C_{j-1},g\right\}  _{{\footnotesize PB}}\nonumber\\
+\sum_{j=1}^{2n-2}\sum_{k\leq j-2}\left(  -1\right)  ^{j+k}\left\{
f,C_{j}\right\}  _{{\footnotesize PB}}\left\{  C_{1},\cdots,C_{k-1}%
,C_{k+1},\cdots,C_{j-1},C_{j+1},\cdots,C_{2n-2}\right\}  _{{\footnotesize NB}%
}\left\{  C_{k},g\right\}  _{{\footnotesize PB}}\nonumber\\
-\sum_{j=1}^{2n-2}\sum_{k\geq j+2}\left(  -1\right)  ^{j+k}\left\{
f,C_{j}\right\}  _{{\footnotesize PB}}\left\{  C_{1},\cdots,C_{j-1}%
,C_{j+1},\cdots,C_{k-1},C_{k+1},\cdots,C_{2n-2}\right\}  _{{\footnotesize NB}%
}\left\{  C_{k},g\right\}  _{{\footnotesize PB}}\;.
\end{gather}
All the remaining NBs contain only constraints. \ Divide by $\left\{
C_{1},C_{2},\cdots\right\}  _{{\footnotesize NB}}$ to find%
\begin{equation}
\frac{1}{\left\{  C_{1},C_{2},\cdots\right\}  _{{\footnotesize NB}}}\left\{
f,g,C_{1},C_{2},\cdots\right\}  _{{\footnotesize NB}}=\left\{  f,g\right\}
_{{\footnotesize PB}}-\sum_{j,k}\left\{  f,C_{j}\right\}  _{{\footnotesize PB}%
}A_{jk}^{-1}\left\{  C_{k},g\right\}  _{{\footnotesize PB}}\;. \label{NBasDB}%
\end{equation}
Here we have noted $\sqrt{\det A}=\left\{  C_{1},C_{2},\cdots,C_{2n-2}%
\right\}  _{{\footnotesize NB}}$, and we have used the expression for the
inverse of $\left\{  C_{j},C_{k}\right\}  =A_{jk}$ in terms of constraint
minors, written as NBs. \ That is,
\begin{equation}
A_{j-1\;j}^{-1}=-A_{j\;j-1}^{-1}=-\left\{  C_{1},\cdots,C_{j-2},C_{j+1}%
,\cdots,C_{2n-2}\right\}  _{{\footnotesize NB}}/\sqrt{\det A}\;,
\end{equation}%
\begin{align}
A_{j\;k\leq j-2}^{-1}  &  =-\left(  -1\right)  ^{j+k}\left\{  C_{1}%
,\cdots,C_{k-1},C_{k+1},\cdots,C_{j-1},C_{j+1},\cdots,C_{2n-2}\right\}
_{{\footnotesize NB}}/\sqrt{\det A}\;,\nonumber\\
A_{j\;k\geq j+2}^{-1}  &  =\left(  -1\right)  ^{j+k}\left\{  C_{1}%
,\cdots,C_{j-1},C_{j+1},\cdots,C_{k-1},C_{k+1},\cdots,C_{2n-2}\right\}
_{{\footnotesize NB}}/\sqrt{\det A}\;.
\end{align}
(That this is correct as an expression for $A_{jk}^{-1}$ follows from
(\ref{NBRecursion}) or its permutations.) \ The RHS of (\ref{NBasDB})\ is
precisely the classical Dirac bracket. \ So we conclude%
\begin{equation}
\left\{  f,g\right\}  _{{\footnotesize DB}}=\frac{1}{\left\{  C_{1}%
,C_{2},\cdots\right\}  _{{\footnotesize NB}}}\left\{  f,g,C_{1},C_{2}%
,\cdots\right\}  _{{\footnotesize NB}}\;. \label{DBasNB}%
\end{equation}
No muss, and no fuss! \ Note that both left- and right-hand sides are
homogeneous of degree zero in each of the constraints.

This last result may be used to provide an alternate proof that the classical
DB obeys the JI. \ Since the classical NB is trivially a derivation,\ we have%
\begin{align}
&  \left(  \left\{  C_{1},C_{2},\cdots\right\}  _{{\footnotesize NB}}\right)
^{3}\left\{  \left\{  f,g\right\}  _{{\footnotesize DB}},h\right\}
_{{\footnotesize DB}}\nonumber\\
&  =\left(  \left\{  C_{1},C_{2},\cdots\right\}  _{{\footnotesize NB}}\right)
^{2}\left\{  \frac{1}{\left\{  C_{1},C_{2},\cdots\right\}
_{{\footnotesize NB}}}\left\{  f,g,C_{1},C_{2},\cdots\right\}
_{{\footnotesize NB}},h,C_{1},C_{2},\cdots\right\}  _{{\footnotesize NB}%
}\nonumber\\
&  =\left\{  C_{1},C_{2},\cdots\right\}  _{{\footnotesize NB}}\;\left\{
\left\{  f,g,C_{1},C_{2},\cdots\right\}  _{{\footnotesize NB}},h,C_{1}%
,C_{2},\cdots\right\}  _{{\footnotesize NB}}\nonumber\\
&  -\left\{  f,g,C_{1},C_{2},\cdots\right\}  _{{\footnotesize NB}}\;\left\{
\left\{  C_{1},C_{2},\cdots\right\}  _{{\footnotesize NB}},h,C_{1}%
,C_{2},\cdots\right\}  _{{\footnotesize NB}}\;.
\end{align}
Upon adding the cycled terms and using the generalization of the fundamental
identity to arbitrary rank CNBs, (\ref{Fundamentalest}), it is now
straightforward to show
\begin{equation}
\left(  \left\{  C_{1},C_{2},\cdots\right\}  _{{\footnotesize NB}}\right)
^{3}\left(  \left\{  \left\{  f,g\right\}  _{{\footnotesize DB}},h\right\}
_{{\footnotesize DB}}+\left\{  \left\{  g,h\right\}  _{{\footnotesize DB}%
},f\right\}  _{{\footnotesize DB}}+\left\{  \left\{  h,f\right\}
_{{\footnotesize DB}},g\right\}  _{{\footnotesize DB}}\right)  =0\;.
\end{equation}
Hence the result (\ref{DBDerivation}). \ 

This derivation of the Jacobi Identity for DBs should be compared with that
for PBs. \ In particular, it is well-known that the simplest conceptual way to
prove the PB JI is as a classical limit ($\hbar\rightarrow0$) of the JI for
quantum commutators, the latter commutator JI being nothing but an encoding of
associativity of trilinear operator products on Hilbert space. \ It might be
asked whether (\ref{DBDerivation}) also follows as a classical limit of a
quantum construction encoding nothing but operator associativity. \ The answer
would be in the affirmative, as the proof we have just sketched using NBs does
indeed follow as a classical limit of an operator statement. \ The identity
(\ref{Fundamentalest}) actually results from the $\hbar\rightarrow0$ limit of
an encoding of nothing but multilinear operator associativity. \ To fully
appreciate this, we must consider the quantization of NBs.

\section{Quantum Theory}

\paragraph{\underline{Definition of QNBs}}

Define the quantum Nambu bracket (QNB) as a fully antisymmetrized multilinear
sum of operator products in an associative enveloping algebra,
\begin{equation}
\left[  A_{1},A_{2},\cdots,A_{k}\right]  \equiv\sum_{\substack{\text{all
}k!\text{\ perms }\left\{  \sigma_{1},\sigma_{2},\cdots,\sigma_{k}\right\}
\\\text{of the indices }\left\{  1,2,\cdots,k\right\}  }}\operatorname{sgn}%
\left(  \sigma\right)  \,A_{\sigma_{1}}A_{\sigma_{2}}\cdots A_{\sigma_{k}}\;,
\end{equation}
where $\operatorname{sgn}\left(  \sigma\right)  =\left(  -1\right)
^{\pi\left(  \sigma\right)  }$ with $\pi\left(  \sigma\right)  $\ the parity
of the permutation $\left\{  \sigma_{1},\sigma_{2},\cdots,\sigma_{k}\right\}
$. \ This definition is also due to Nambu \cite{Nambu}, although such
structures have independently appeared in the mathematical literature
\cite{Kurosh}.

\paragraph{\underline{Recursion relations}}

There are various ways to obtain QNBs recursively, from products involving
fewer operators. \ For example, a QNB involving $k$ operators has both left-
and right-sided resolutions of single operators multiplying QNBs of $k-1$
operators.%
\begin{align}
\left[  A_{1},A_{2},\cdots,A_{k}\right]   &  =\sum_{k!\;\text{perms}\;\sigma
}\frac{\operatorname{sgn}\left(  \sigma\right)  }{\left(  k-1\right)
!}\,A_{\sigma_{1}}\left[  A_{\sigma_{2}},\cdots,A_{\sigma_{k}}\right]
\nonumber\\
&  =\sum_{k!\;\text{perms}\;\sigma}\frac{\operatorname{sgn}\left(
\sigma\right)  }{\left(  k-1\right)  !}\,\left[  A_{\sigma_{1}},\cdots
,A_{\sigma_{k-1}}\right]  A_{\sigma_{k}}\;.
\end{align}
On the RHS there are actually only $k$ distinct products of single elements
with $\left(  k-1\right)  $-brackets, each such product having a net
coefficient $\pm1$. \ The denominator compensates for replication of these
products in the sum over permutations. \ (We leave it as an elementary
exercise for the reader to prove this result.)

For example, the 2-bracket is obviously just the commutator $\left[
A,B\right]  =AB-BA$,\ while the 3-bracket may be written in either of two
convenient ways,%
\begin{align}
\left[  A,B,C\right]   &  =A\left[  B,C\right]  +B\left[  C,A\right]
+C\left[  A,B\right] \nonumber\\
&  =\left[  A,B\right]  C+\left[  B,C\right]  A+\left[  C,A\right]  B\;.
\end{align}
Summing these two RHS lines gives anticommutators containing commutators on
the RHS.%
\begin{equation}
2\times\left[  A,B,C\right]  =\left\{  A,\left[  B,C\right]  \right\}
+\left\{  B,\left[  C,A\right]  \right\}  +\left\{  C,\left[  A,B\right]
\right\}  \;.
\end{equation}
The last expression is to be contrasted to the Jacobi identity obtained by
taking the \emph{difference} of the two RHS lines.
\begin{equation}
0=\left[  A,\left[  B,C\right]  \right]  +\left[  B,\left[  C,A\right]
\right]  +\left[  C,\left[  A,B\right]  \right]  \;.
\end{equation}
Similarly for the 4-bracket,%
\begin{align}
\left[  A,B,C,D\right]   &  =A\left[  B,C,D\right]  -B\left[  C,D,A\right]
+C\left[  D,A,B\right]  -D\left[  A,B,C\right] \nonumber\\
&  =-\left[  B,C,D\right]  A+\left[  C,D,A\right]  B-\left[  D,A,B\right]
C+\left[  A,B,C\right]  D\;.
\end{align}
Summing these two lines gives%
\begin{equation}
2\times\left[  A,B,C,D\right]  =\left[  A,\left[  B,C,D\right]  \right]
-\left[  B,\left[  C,D,A\right]  \right]  +\left[  C,\left[  D,A,B\right]
\right]  -\left[  D,\left[  A,B,C\right]  \right]  \;,
\end{equation}
while taking the difference gives%
\begin{equation}
0=\left\{  A,\left[  B,C,D\right]  \right\}  -\left\{  B,\left[  C,D,A\right]
\right\}  +\left\{  C,\left[  D,A,B\right]  \right\}  -\left\{  D,\left[
A,B,C\right]  \right\}  \;.
\end{equation}
There may be some temptation to think of the last of these as something like a
generalization of the Jacobi identity, and in principle, it is, but in a a
crucially limited way, so that temptation should be checked. The more
appropriate and complete generalization of the Jacobi identity is given
systematically below.

\paragraph{\underline{Jordan products}}

Define a fully symmetrized, generalized Jordan operator product, or GJP,%
\begin{equation}
\left\{  A_{1},A_{2},\cdots,A_{k}\right\}  \equiv\sum_{\substack{\text{all
}k!\text{\ perms }\left\{  \sigma_{1},\sigma_{2},\cdots,\sigma_{k}\right\}
\\\text{of the indices }\left\{  1,2,\cdots,k\right\}  }}A_{\sigma_{1}%
}A_{\sigma_{2}}\cdots A_{\sigma_{k}}\;,
\end{equation}
as introduced, in the bilinear form at least, by Pascual Jordan to render
non-abelian algebras into abelian algebras at the expense of
non-associativity. \ The generalization to multi-linears was suggested by
Kurosh \cite{Kurosh}, but the idea was not used in any previous physical
application, as far as we know. \ 

A GJP also has left- and right-sided recursions,
\begin{align}
\left\{  A_{1},A_{2},\cdots,A_{k}\right\}   &  =\sum_{k!\;\text{perms}%
\;\sigma}\frac{1}{\left(  k-1\right)  !}\,A_{\sigma_{1}}\left\{  A_{\sigma
_{2}},A_{\sigma_{3}},\cdots,A_{\sigma_{k}}\right\} \nonumber\\
&  =\sum_{k!\;\text{perms}\;\sigma}\frac{1}{\left(  k-1\right)  !}\,\left\{
A_{\sigma_{2}},A_{\sigma_{3}},\cdots,A_{\sigma_{k-1}}\right\}  A_{\sigma_{k}%
}\;.
\end{align}
On the RHS there are again only $k$ distinct products of single elements with
$\left(  k-1\right)  $-GJPs, each such product having a net coefficient $+1$.
\ The denominator again compensates for replication of these products in the
sum over permutations. \ (We leave it as another elementary exercise for the
reader to prove this result.)

For example, a Jordan 2-product is obviously just an anticommutator $\left\{
A,B\right\}  =AB+BA$, while a 3-product is given by%
\begin{align}
\left\{  A,B,C\right\}   &  =\left\{  A,B\right\}  C+\left\{  A,C\right\}
B+\left\{  B,C\right\}  A\nonumber\\
&  =A\left\{  B,C\right\}  +B\left\{  A,C\right\}  +C\left\{  A,B\right\}  \;.
\end{align}
Equivalently, taking sums and differences, we obtain%
\begin{equation}
2\times\left\{  A,B,C\right\}  =\left\{  A,\left\{  B,C\right\}  \right\}
+\left\{  B,\left\{  A,C\right\}  \right\}  +\left\{  C,\left\{  A,B\right\}
\right\}  \;,
\end{equation}
as well as the companion of the Jacobi identity often encountered in
super-algebras,%
\begin{equation}
0=\left[  A,\left\{  B,C\right\}  \right]  +\left[  B,\left\{  A,C\right\}
\right]  +\left[  C,\left\{  A,B\right\}  \right]  \;.
\end{equation}
Similarly for the 4-product,%
\begin{align}
\left\{  A,B,C,D\right\}   &  =A\left\{  B,C,D\right\}  +B\left\{
C,D,A\right\}  +C\left\{  D,A,B\right\}  +D\left\{  A,B,C\right\} \nonumber\\
&  =\left\{  A,B,C\right\}  D+\left\{  B,C,D\right\}  A+\left\{
C,D,A\right\}  B+\left\{  D,A,B\right\}  C\;.
\end{align}
Summing gives%
\begin{equation}
2\times\left\{  A,B,C,D\right\}  =\left\{  A,\left\{  B,C,D\right\}  \right\}
+\left\{  B,\left\{  C,D,A\right\}  \right\}  +\left\{  C,\left\{
D,A,B\right\}  \right\}  +\left\{  D,\left\{  A,B,C\right\}  \right\}  \;,
\end{equation}
while subtracting gives%
\begin{equation}
0=\left[  A,\left\{  B,C,D\right\}  \right]  +\left[  B,\left\{
C,D,A\right\}  \right]  +\left[  C,\left\{  D,A,B\right\}  \right]  +\left[
D,\left\{  A,B,C\right\}  \right]  \;.
\end{equation}
Again the reader is warned off the temptation to think of the last of these as
a bona fide generalization of the super-Jacobi identity. \ It is a valid
identity of course, following from nothing but associativity, but there is a
superior and complete set of identities to be given later.

\paragraph{\underline{(Anti)Commutator resolutions}}

As in the classical case, it is always possible to resolve even rank brackets
into sums of commutator products, very usefully. \ For example,
\begin{align}
\left[  A,B,C,D\right]   &  =\left[  A,B\right]  \left[  C,D\right]  -\left[
A,C\right]  \left[  B,D\right]  -\left[  A,D\right]  \left[  C,B\right]
\nonumber\\
&  +\left[  C,D\right]  \left[  A,B\right]  -\left[  B,D\right]  \left[
A,C\right]  -\left[  C,B\right]  \left[  A,D\right]  \;.
\end{align}
An arbitrary even bracket of rank $2n$ breaks up into $\left(  2n\right)
!/\left(  2^{n}\right)  =n!\left(  2n-1\right)  !!$ such products. \ Another
way to say this is that even QNBs can be written in terms of GJPs of
commutators. \ The general result is%
\begin{equation}
\left[  A_{1},\cdots,A_{2n}\right]  =\sum_{\left(  2n\right)  !\;\text{perms}%
\;\sigma}\frac{\operatorname{sgn}\left(  \sigma\right)  }{2^{n}n!}\,\left\{
\left[  A_{\sigma_{1}},A_{\sigma_{2}}\right]  ,\cdots,\left[  A_{\sigma
_{2n-1}},A_{\sigma_{2n}}\right]  \right\}  \;.
\end{equation}
An even GJP also resolves into symmetrized products of anticommutators.%
\begin{equation}
\left\{  A_{1},\cdots,A_{2n}\right\}  =\sum_{\left(  2n\right)
!\;\text{perms}\;\sigma}\frac{1}{2^{n}n!}\,\left\{  \left\{  A_{\sigma_{1}%
},A_{\sigma_{2}}\right\}  ,\cdots,\left\{  A_{\sigma_{2n-1}},A_{\sigma_{2n}%
}\right\}  \right\}  \;.
\end{equation}
As in the classical bracket formalism, the proofs of these relations are
elementary. \ Both left- and right-hand sides of the expressions are sums of
$2n$-th degree monomials linear in each of the $A$s. \ Both sides are either
totally antisymmetric, in the case of QNBs, or totally symmetric, in the case
of GJPs, under permutations of the $A$s. \ Thus the two sides must be
proportional. \ The only issue left is the constant of proportionality. \ This
is easily determined to be $1$ just by comparing the coefficients of any given
term appearing on both sides of the equation, e.g. $A_{1}A_{2}\cdots
A_{2N-1}A_{2N}$. \ 

\paragraph{\underline{The classical limit}}

Since Poisson brackets are straightforward classical limits of commutators,
\begin{equation}
\lim_{\hbar\rightarrow0}\left(  \frac{1}{i\hbar}\right)  \left[  A,B\right]
=\left\{  A,B\right\}  _{{\footnotesize PB}}\;,
\end{equation}
it follows that the commutator resolution of all even QNBs directly specifies
their classical limit. \ (For a detailed approach to the classical limit,
including sub-dominant terms of higher order in $\hbar$, see, e.g., the Moyal
Bracket discussion in \cite{ZC}.)

For example, from%
\begin{equation}
\left[  A,B,C,D\right]  =\left\{  \left[  A,B\right]  ,\left[  C,D\right]
\right\}  -\left\{  \left[  A,C\right]  ,\left[  B,D\right]  \right\}
-\left\{  \left[  A,D\right]  ,\left[  C,B\right]  \right\}  \;,
\end{equation}
with due attention to factors of $2$, the classical limit emerges as%
\begin{align}
&  \frac{1}{2}\times\lim_{\hbar\rightarrow0}\left(  \frac{1}{i\hbar}\right)
^{2}\left[  A,B,C,D\right] \nonumber\\
&  =\left\{  A,B\right\}  _{{\footnotesize PB}}\left\{  C,D\right\}
_{{\footnotesize PB}}-\left\{  A,C\right\}  _{{\footnotesize PB}}\left\{
B,D\right\}  _{{\footnotesize PB}}-\left\{  A,D\right\}  _{{\footnotesize PB}%
}\left\{  C,B\right\}  _{{\footnotesize PB}}\nonumber\\
&  =\left\{  A,B,C,D\right\}  _{{\footnotesize NB}}\;.
\end{align}
And so it goes with all other even rank Nambu brackets. \ For a $2n$-bracket,
one sees that
\begin{align}
&  \frac{1}{n!}\times\lim_{\hbar\rightarrow0}\left(  \frac{1}{i\hbar}\right)
^{n}\left[  A_{1},\cdots,A_{2n}\right] \nonumber\\
&  =\sum_{\left(  2n\right)  !\text{\ perms }\sigma}\frac{\operatorname{sgn}%
\left(  \sigma\right)  }{2^{n}n!}\,\left\{  A_{\sigma_{1}},A_{\sigma_{2}%
}\right\}  _{{\footnotesize PB}}\left\{  A_{\sigma_{3}},A_{\sigma_{4}%
}\right\}  _{{\footnotesize PB}}\cdots\left\{  A_{\sigma_{2n-1}}%
,A_{\sigma_{2n}}\right\}  _{{\footnotesize PB}}\nonumber\\
&  =\left\{  A_{1},\cdots,A_{2n}\right\}  _{{\footnotesize NB}}\;.
\end{align}
This is another way to establish that there are indeed $\left(  2n-1\right)
!! $ independent products of $n$ Poisson brackets summing up to give the PB
resolution of the classical Nambu $2n$-bracket. \ Once again due attention
must be given to a critical additional factor of $n!$ (as in the denominator
on the LHS) since the GJPs on the RHS of the commutator resolution will, in
the classical limit, always replicate the same classical product $n!$ times.

\paragraph{\underline{The Leibniz rule failure and derivators}}

Define the \emph{derivator} to measure the failure of the simplest Leibniz
rule for QNBs,
\begin{gather}
^{k+1}\mathbf{\Delta}_{\mathbf{B}}\left(  A,\mathcal{A}\right)  \equiv\left(
A,\mathcal{A}\,|\,B_{1},\cdots,B_{k}\right) \nonumber\\
\equiv\left[  \,A\mathcal{A\,},B_{1},\cdots,B_{k}\right]  -A\left[
\mathcal{A},B_{1},\cdots,B_{k}\right]  -\left[  A,B_{1},\cdots,B_{k}\right]
\mathcal{A}\;.
\end{gather}
The first term on the RHS is a $\left(  k+1\right)  $-bracket acting on just
the product of $A$\ and $\mathcal{A}$, the order of the bracket being evident
in the pre-superscript of the $\mathbf{\Delta}_{\mathbf{B}}$ notation. \ This
reads in an obvious way. \ For instance, $^{4}\mathbf{\Delta}_{\mathbf{B}}$ is
a ``4-delta of $B$s''. \ That notation also emphasizes that the $B$s act
\emph{on} the pair of $A$s. \ The second notation makes explicit all the $B$s
and is useful for computer code. \ 

Any $\mathbf{\Delta}_{\mathbf{B}}$ acts on all pairs of elements in the
enveloping algebra $\mathfrak{A}$ to produce another element in $\mathfrak{A}
$.%
\[
\mathbf{\Delta}_{\mathbf{B}}:\mathfrak{A}\times\mathfrak{A}\longmapsto
\mathfrak{A}\;.
\]
When $\mathbf{\Delta}_{\mathbf{B}}$ does not vanish the corresponding bracket
with the $B$s does not define a derivation on $\mathfrak{A}$. \ The derivator
$\mathbf{\Delta}_{\mathbf{B}}\left(  A,\mathcal{A}\right)  $ is linear in both
$A$ and $\mathcal{A},$ as well as linear in each of the $B$s.

Less trivially, from explicit calculations, we find inhomogeneous recursion
relations for these derivators.%
\begin{gather}
\left(  A,\mathcal{A}\,|\,B_{1},\cdots,B_{k}\right) \nonumber\\
=\sum_{k!\text{\ perms }\sigma}\frac{\frac{1}{2}\operatorname{sgn}\left(
\sigma\right)  }{\left(  k-1\right)  !}\left(  \left(  A,\mathcal{A}%
\,|\,B_{\sigma_{1}},\cdots,B_{\sigma_{k-1}}\right)  B_{\sigma_{k}}+\left(
-1\right)  ^{k}B_{\sigma_{k}}\left(  A,\mathcal{A}\,|\,B_{\sigma_{1}}%
,\cdots,B_{\sigma_{k-1}}\right)  \right) \nonumber\\
+\sum_{k!\text{\ perms }\sigma}\frac{\frac{1}{2}\operatorname{sgn}\left(
\sigma\right)  }{\left(  k-1\right)  !}\left(  \left[  A,B_{\sigma_{k}%
}\right]  \left[  B_{\sigma_{1}},\cdots,B_{\sigma_{k-1}},\mathcal{A}\right]
-\left[  A,B_{\sigma_{1}},\cdots,B_{\sigma_{k-1}}\right]  \left[
B_{\sigma_{k}},\mathcal{A}\right]  \right) \nonumber\\
+\frac{\left(  -1\right)  ^{k+1}-1}{2}\,A\left[  B_{1},\cdots,B_{k}\right]
\mathcal{A}\;.
\end{gather}
Alternatively, we may write this so as to emphasize the number of distinct
terms on the RHS and distinguish between the even and odd bracket cases. \ The
first two terms under the sum on the RHS give a commutator/anticommutator for
$k$ odd/even, and the last term is absent for $k$ odd, while for $k$ even it
may be viewed as a type of obstruction in the recursion relation for the odd
quantum bracket.

The obstruction is clarified if we specialize to $n=1$, i.e. the 3-bracket
case. \ Since commutators are always derivations, one has $^{2}\mathbf{\Delta
}_{B}\left(  A,\mathcal{A}\right)  =0$, and the first RHS line vanishes\ for
the $^{3}\mathbf{\Delta}_{\mathbf{B}}\left(  A,\mathcal{A}\right)  $ case
above. \ So we have just
\begin{equation}
\left(  A,\mathcal{A}\,|\,B_{1},B_{2}\right)  =\left[  A,B_{2}\right]  \left[
B_{1},\mathcal{A}\right]  -\left[  A,B_{1}\right]  \left[  B_{2}%
,\mathcal{A}\right]  -A\left[  B_{1},B_{2}\right]  \mathcal{A}\;.
\end{equation}
The first two terms on the RHS are $O\left(  \hbar^{2}\right)  $\ while the
last is $O\left(  \hbar\right)  $. \ It is precisely this last term which was
responsible for some of Nambu's misgivings concerning his quantum 3-bracket.
\ In particular, even in the extreme case when both $A$ and $\mathcal{A}$
commute with the $B$s, $^{3}\mathbf{\Delta}_{\mathbf{B}}\left(  A,\mathcal{A}%
\right)  $ does not vanish: $\left.  \left(  A,\mathcal{A};B_{1},B_{2}\right)
\right|  _{\left[  A,B_{i}\right]  =0=\left[  \mathcal{A},B_{i}\right]
}=-A\mathcal{A}\,\left[  B_{1},B_{2}\right]  $. \ By contrast, for the even
(2n+2)-bracket, all terms on the RHS are generically of the same order,
$O\left(  \hbar^{n+1}\right)  $. \ In terms of combinatorics, this seems to be
the only feature for the simple, possibly failed, Leibniz rule that
distinguishes between even and odd brackets.

\paragraph{\underline{Generalized Jacobi identities and QFIs}}

We previously pointed out some elementary identities involving QNBs which are
suggestive of generalizations of the Jacobi identity for commutators. \ Those
particular identities, while true, were not designated as ``generalized Jacobi
identities'', for the simple fact that they do \emph{not} involve the case
where QNBs of a given rank act on QNBs of the same rank. \ Here we explore QNB
identities of the latter type. \ There are indeed acceptable generalizations
of the usual commutators-acting-on-commutators Jacobi identity (i.e. quantum
2-brackets acting on quantum 2-brackets), and these generalizations are indeed
valid for \emph{all} higher rank QNBs (i.e. quantum n-brackets acting on
quantum n-brackets). \ However, there is an essential distinction to be drawn
between the even and odd quantum bracket cases.

It is important to note that, historically, there have been some incorrect
guesses and false starts in this direction that originated from the so-called
fundamental identity obeyed by classical Nambu brackets (see \cite{CQNM} for
the irrelevant literature). \ The correct generalizations of the Jacobi
identities which \emph{do} encode associativity were found independently by
groups of mathematicians\ and physicists \cite{Hanlon,Azcarraga}.
\ Interestingly, both groups were studying cohomology questions, so perhaps it
is not surprising that they arrived at the same result. \ The acceptable
generalization of the Jacobi identity that was found is satisfied by all QNBs,
although for odd QNBs there is a significant difference in the form of the
final result: \ It contains an ``inhomogeneity''. \ The correct generalization
is obtained just by totally antisymmetrizing the action of one n-bracket on
the other. \ Effectively, this amounts to antisymmetrizing the form of the
classical FI over all permutations of the $A$s and $B$s, including \emph{all}
exchanges\emph{\ }of $A$s \emph{with} $B$s. \ To describe some of the results
before giving them in detail:

\begin{center}
\textit{The totally antisymmetrized action of odd }$n$\textit{\ QNBs on other
odd }$n$\textit{\ QNBs }

\textit{results in }$\left(  2n-1\right)  $\textit{-brackets.}

\textit{The totally antisymmetrized action of even }$n$\textit{\ QNBs on other
even }$n$\textit{\ QNBs }

\textit{results in zero.}
\end{center}

\noindent More precisely, the generalized Jacobi identities (GJIs) for
arbitrary $n$-brackets follow from totally antisymmetrizing the action of any
bracket on any other through use of the so-called ``shifting bracket
argument'' \cite{CQNM}. \ That argument actually leads to a larger set of
results which we summarize here, calling them the \emph{quantum Jacobi
identities}, or QJIs. \ The GJIs are special cases of QJIs for $k=n-1$.

\paragraph{\underline{QJI for QNBs}}%

\begin{align}
&  \sum_{\left(  n+k\right)  !\;\text{perms }\sigma}\operatorname{sgn}\left(
\sigma\right)  \left[  \left[  A_{\sigma_{1}},\cdots,A_{\sigma_{n}}\right]
,A_{\sigma_{n+1}},\cdots,A_{\sigma_{n+k}}\right] \nonumber\\
&  =\left[  A_{1},\cdots,A_{n+k}\right]  \times n!k!\times\left\{
\begin{array}
[c]{c}%
\left(  k+1\right)  \text{ \ \ if }n\text{\ is odd}\\
\frac{1}{2}\left(  1+\left(  -1\right)  ^{k}\right)  \text{ \ \ if
}n\text{\ is even}%
\end{array}
\right.  \;. \label{QJI}%
\end{align}
This result is proven just by computing the coefficient of any selected
monomial, e.g. $A_{1}\cdots A_{n+k}$. \ This is the quantum identity that most
closely corresponds to the general classical result (\ref{Fundamentalest}).
\ While that classical identity holds without requiring antisymmetrization
over exchanges of $A$s and $B$s, in contrast the quantum identity \emph{must}
be totally antisymmetrized if it is to be a consequence of only the
associativity of the underlying algebra of Hilbert space operators. \ Note
that the $n!k!$ on the RHS of (\ref{QJI}) may be replaced by just $1$ if we
sum \emph{only} over permutations in which the $A_{i\leq n} $ are interchanged
with the $A_{i>n}$ in $\left[  \left[  A_{\sigma_{1}},\cdots,A_{\sigma_{n}%
}\right]  ,A_{\sigma_{n+1}},\cdots,A_{\sigma_{n+k}}\right]  $, and ignore all
permutations of the $A_{1},A_{2},\cdots,A_{n}$ among themselves, and of the
$A_{n+1},\cdots,A_{n+k}$ among themselves.

There is an important specialization of the QJI result: \ For even $n$ and odd
$k,$%
\begin{equation}
\sum_{\left(  n+k\right)  !\;\text{perms}\;\sigma}\operatorname{sgn}\left(
\sigma\right)  \left[  \left[  A_{\sigma_{1}},\cdots,A_{\sigma_{n}}\right]
,A_{\sigma_{n+1}},\cdots,A_{\sigma_{n+k}}\right]  =0\;.
\end{equation}
In particular, when $k=n-1,$ for $n$ even, the vanishing RHS obtains. \ All
other $n$-not-even and/or $k$-not-odd cases of the QJI have the $\left[
A_{1},\cdots,A_{n+k}\right]  $ inhomogeneity on the RHS. \ The classical limit
of this last identity is immediate.%
\begin{equation}
\sum_{\left(  n+k\right)  !\;\text{perms}\;\sigma}\operatorname{sgn}\left(
\sigma\right)  \left\{  \left\{  A_{\sigma_{1}},\cdots,A_{\sigma_{n}}\right\}
_{NB},A_{\sigma_{n+1}},\cdots,A_{\sigma_{n+k}}\right\}  _{NB}=0\;.
\end{equation}
For example, when the outside bracket is just a PB, this implies for odd $k$%
\begin{equation}
\varepsilon_{a_{1}\cdots a_{k}}\left\{  \left\{  B_{a_{1}},\cdots,B_{a_{k-1}%
}\right\}  ,B_{a_{k}}\right\}  =0\;,
\end{equation}
where $\varepsilon_{a_{1}\cdots a_{k}}$ is the totally antisymmetric
Levi-Civita symbol, and all $a_{i}$ are summed from $1$ to $k$. \ This
classical identity is needed to obtain the form of the terms in the last line
of (\ref{Fundamentalest}), and hence to prove the DB JI.

The QJI also permits us to give the correct operator form of the so-called
fundamental identities valid for all QNBs. \ We accordingly call these
\emph{quantum fundamental identities}, or QFIs, and present them in their
general form.

\paragraph{\underline{QFI for QNBs}}%

\begin{align}
&  \sum_{\left(  n+k\right)  !\;\text{perms}\;\sigma}\operatorname{sgn}\left(
\sigma\right)  {\Large (}\left[  \left[  A_{\sigma_{1}},\cdots,A_{\sigma_{n}%
}\right]  ,A_{\sigma_{n+1}},\cdots,A_{\sigma_{n+k}}\right] \nonumber\\
&  -\sum_{j=1}^{n}\left[  A_{\sigma_{1}},\cdots,\left[  A_{\sigma_{j}%
},A_{\sigma_{n+1}},\cdots,A_{\sigma_{n+k}}\right]  ,\cdots,A_{\sigma_{n}%
}\right]  {\Large )}\nonumber\\
&  =\left[  A_{1},\cdots,A_{n+k}\right]  \times n!k!\times\left\{
\begin{array}
[c]{c}%
0\text{ \ \ if }k\text{\ is odd}\\
\left(  1-n\right)  \left(  k+1\right)  \text{ \ \ if }k\text{\ is even and
}n\text{\ is odd}\\
\left(  1-n\left(  k+1\right)  \right)  \text{ \ \ if }k\text{\ is even and
}n\text{\ is even}%
\end{array}
\right.  \;. \label{QFI}%
\end{align}
Aside from the trivial case of $n=1$, the only way the RHS vanishes without
conditions on the full $\left(  n+k\right)  $-bracket is when $k$ is odd.
\ All $n>1$, even $k$ result in the $\left[  A_{1},\cdots,A_{n+k}\right]  $
inhomogeneity on the RHS.

Partial antisymmetrizations of the individual terms in the general QFI may
also be entertained. \ The result is to find more complicated inhomogeneities,
and does not seem to be very informative. \ At best these partial
antisymmetrizations show in a supplemental way how the fully antisymmetrized
results are obtained. \ In certain isolated, special cases (cf. the $su\left(
2\right)  $ example of the next section, for which $k=3$), the bracket effects
of select $B$s can act as a derivation (essentially because the k-bracket is
equivalent, in its effects, to a commutator). \ If that is the case, then the
quantum version of the classical FI (\ref{Fundamental}) \emph{trivially}
holds. \ It is also possible in principle for that simple identity to hold,
again in very special situations, if the quantum bracket is not a derivation,
through various cancellations among terms. \ As an aid to finding such
peculiar situations, it is useful to resolve the quantum correspondents of the
terms in the classical FI into the derivators introduced previously. \ From
the definition of $\left[  A_{1},\cdots,A_{n}\right]  $, and some
straightforward manipulations, we find
\begin{gather}
\left[  \left[  A_{1},\cdots,A_{n}\right]  ,\mathbf{B}\right]  -\sum_{j=1}%
^{n}\left[  A_{1},\cdots,\left[  A_{j},\mathbf{B}\right]  ,\cdots,A_{n}\right]
\nonumber\\
=\sum_{n!\text{\ perms }\sigma}\operatorname{sgn}\left(  \sigma\right)
{\Large (}\tfrac{1}{\left(  n-1\right)  !}\left(  A_{\sigma_{1}},\left[
A_{\sigma_{2}},\cdots,A_{\sigma_{n}}\right]  \,|\,\mathbf{B}\right)
+\tfrac{1}{\left(  n-2\right)  !}A_{\sigma_{1}}\left(  A_{\sigma_{2}},\left[
A_{\sigma_{3}},\cdots,A_{\sigma_{n}}\right]  \,|\,\mathbf{B}\right)
\nonumber\\
+\tfrac{1}{2!\left(  n-3\right)  !}\left[  A_{\sigma_{1}},A_{\sigma_{2}%
}\right]  \left(  A_{\sigma_{3}},\left[  A_{\sigma_{4}},\cdots,A_{\sigma_{n}%
}\right]  \,|\,\mathbf{B}\right) \nonumber\\
+\cdots+\tfrac{1}{\left(  n-1\right)  !}\left[  A_{\sigma_{1}},A_{\sigma_{2}%
},\cdots,A_{\sigma_{n-2}}\right]  \left(  A_{\sigma_{n-1}},A_{\sigma_{n}%
}\,|\,\mathbf{B}\right)  {\Large )\;,}%
\end{gather}
with the abbreviation $\mathbf{B=}B_{1},\cdots,B_{k}$. \ The terms on the RHS
are a sum over $j=1,\cdots,n-1$ of derivators between solitary $A$s (i.e.
$1$-brackets) and various $\left(  n-j\right)  $-brackets, left-multiplied by
complementary rank $\left(  j-1\right)  $-brackets. \ (There is a similar
identity that involves right-multiplication by the complementary brackets.)
\ For example, suppose $n=2$. \ Then we have for any number of $B$s%
\begin{equation}
\left[  \left[  A_{1},A_{2}\right]  ,\mathbf{B}\right]  -\left[  \left[
A_{1},\mathbf{B}\right]  ,A_{2}\right]  -\left[  A_{1},\left[  A_{2}%
,\mathbf{B}\right]  \right]  =\left(  A_{1},A_{2}\,|\,\mathbf{B}\right)
-\left(  A_{2},A_{1}\,|\,\mathbf{B}\right)  \;.
\end{equation}
In principle, this can vanish, even when the action of the $B$s is not a
derivation, if the k-derivator is symmetric in the first two arguments. \ That
is, if $\left(  A_{1},A_{2}\,|\,B_{1},\cdots,B_{k}\right)  =\tfrac{1}%
{2}\left(  A_{1},A_{2}\,|\,B_{1},\cdots,B_{k}\right)  +\tfrac{1}{2}\left(
A_{2},A_{1}\,|\,B_{1},\cdots,B_{k}\right)  \;$. \ However, we have \emph{not}
found an interesting (nontrivial) physical example where this is the case.

\section{Quantum entwinement of Dirac Brackets}

From the commutator resolution of the QNB,%
\begin{multline}
\left[  f,g,C_{1},C_{2},\cdots\right]  =\left\{  \left[  f,g\right]  ,\left[
C_{1},C_{2}\right]  ,\cdots\right\}  \pm\text{ \ \ permutations of the
}C\text{s}\nonumber\\
-\left\{  \left[  f,C_{1}\right]  ,\left[  g,C_{2}\right]  ,\cdots\right\}
+\left\{  \left[  g,C_{1}\right]  ,\left[  f,C_{2}\right]  ,\cdots\right\}
\pm\text{ \ \ permutations of the }C\text{s.}%
\end{multline}
Terms in the first line, where the commutator $\left[  f,g\right]  $ is
intact, are the quantum analogues of the first line in the classical relation%
\begin{gather}
\left\{  f,g,C_{1},C_{2},\cdots\right\}  _{{\footnotesize NB}}=\left\{
f,g\right\}  _{{\footnotesize PB}}\left\{  C_{1},C_{2},\cdots\right\}
_{{\footnotesize NB}}\nonumber\\
+\sum_{j,k}\left(  -1\right)  ^{j+k}\left\{  f,C_{j}\right\}
_{{\footnotesize PB}}\left\{  C_{1},\cdots,C_{k-1},C_{k+1},\cdots
,C_{j-1},C_{j+1},\cdots\right\}  _{{\footnotesize NB}}\left\{  C_{k}%
,g\right\}  _{{\footnotesize PB}}\;,
\end{gather}
whereas terms in the commutator resolution where $f$ and $g$ appear in
different commutators, as in the second line above, are the quantum analogues
of the second line in the classical relation. \ Unfortunately, the coefficient
of $\left[  f,g\right]  $ cannot be trivially divided out of the commutator
resolution as it could in the classical relation. \ That is, generalized
special Jordan algebras are not division rings (see the Appendix, \cite{CQNM},
for an exemplary discussion). \ 

However, if the Hilbert space is suitably partitioned into invariant sectors
using projectors, with%
\begin{equation}
\left[  f,g\right]  =\sum_{j,k}\mathbb{P}_{j}\left(  \left[  f,g\right]
\right)  _{jk}\mathbb{P}_{k}\;,\;\;\;\mathbb{P}_{k}^{2}=\mathbb{P}%
_{k}\;,\;\;\;\mathbb{P}_{j}\mathbb{P}_{k}=0\;,\;\;\;j\neq k\;.
\end{equation}
then the cumulative effect of the $\left[  C_{i},C_{j}\right]  $ terms can
possibly be diagonalized, \
\begin{equation}
\sigma\times\left(  \left[  f,g\right]  \right)  _{\sigma}=\left\{  \left(
\left[  f,g\right]  \right)  _{\sigma},\left[  C_{1},C_{2}\right]
,\cdots\right\}  \pm\ \ \text{permutations,}%
\end{equation}
and hence for non-vanishing $\sigma$ the effects of entwining $\left[
f,g\right]  $ within the constraint commutators can be divided out, sector by
sector. \ On each such invariant sector, an effective Dirac bracket can then
be computed. \ 

The result is \textbf{not} a derivation, on the full Hilbert space, but acts
like a derivation on each invariant sector. \ This is the best we can do to
define QDBs in terms of QNBs, in general. \ For further discussion of the use
of such projections as well as the physics coded in QNBs for particular
examples, see \cite{CQNM}, and also \cite{ZC} in these proceedings. \ Another
recent proposal to implement constraints using projection operators, which
came to our attention while writing up this contribution, is to be found in
\cite{Balatin}.

\paragraph{Acknowledgements:}

We thank Y Nambu, D Fairlie, Y Nutku, and J de Azc\'{a}rraga for useful comments.

\end{document}